\documentclass[11pt,reqno]{amsart}
\usepackage{amsmath,amssymb,mathrsfs,needspace}
\usepackage[colorlinks=true,linkcolor=blue,citecolor=blue,urlcolor=blue]{hyperref}
\allowdisplaybreaks
\newtheorem{theorem}{Theorem}[section]
\newtheorem{lemma}[theorem]{Lemma}
\newtheorem{proposition}[theorem]{Proposition}
\newtheorem{corollary}[theorem]{Corollary}

\newcommand{\C}{\mathbb C}
\newcommand{\tr}{\operatorname{tr}}

\newcommand{\NC}{\operatorname{NC}}
\newcommand{\Fock}{\mathfrak F}
\newcommand{\id}{\mathrm{id}}

\begin{document}
\title{A Reduction Bound for Entanglement Detection by Positive Free Maps}
\author[Lee]{Chin-Yen Lee}
\email{109281001@cc.ncu.edu.tw}
\begin{abstract}
Collins, Hayden, and Nechita used free probability to construct families of $k$-positive but not $(k+1)$-positive maps and applied them to entanglement detection. We prove that positive free maps associated with compactly supported freely infinitely divisible distributions are no stronger than the reduction criterion in every ancillary dimension. In particular, these maps cannot detect positive partial transpose entanglement. The proof combines a joint Fock realization with an operator positivity certificate derived from free L\'evy data.
\end{abstract}
\maketitle

%%%%%%%%%%%%%%%%%%%%%%%%%%%%%%%%%%%%%%%%%%%%%%%%%%%%%%%%%%%%%%%%%%%%%%%%%%
%%%%%%%%%%%%%%%%%%%%%%%%%%%%%%%%%%%%%%%%%%%%%%%%%%%%%%%%%%%%%%%%%%%%%%%%%%
%%%
%%% Section 1
%%%
%%%%%%%%%%%%%%%%%%%%%%%%%%%%%%%%%%%%%%%%%%%%%%%%%%%%%%%%%%%%%%%%%%%%%%%%%%
%%%%%%%%%%%%%%%%%%%%%%%%%%%%%%%%%%%%%%%%%%%%%%%%%%%%%%%%%%%%%%%%%%%%%%%%%%
\section{Introduction}\label{sec:intro}

Positive maps provide necessary separability tests for bipartite quantum states: every positive linear map $\Phi:\mathrm{Mat}_n(\C)\to\mathcal A$, with $\mathcal A$ a $C^*$-algebra, satisfies $(\Phi\otimes\id_m)(\rho)\succeq0$ whenever a state $\rho\in\mathrm{Mat}_n(\C)\otimes\mathrm{Mat}_m(\C) \cong\mathrm{Mat}_n(\mathrm{Mat}_m(\C))$ is separable.
A violation of this positivity condition therefore certifies entanglement. 
For each positive map $\Phi$, define its \emph{undetected set} by
\begin{align*}
K_\Phi^{(n,m)}
=\bigl\{
\rho\in\mathrm{Mat}_n(\mathrm{Mat}_m(\C)):\rho\succeq 0,\,
\operatorname{Tr}(\rho)=1,\ 
(\Phi\otimes\id_m)(\rho)\succeq0
\bigr\}.
\end{align*}
Thus, a state is detected by $\Phi$ precisely when
it lies outside $K_\Phi^{(n,m)}$.

Two standard examples \cite{CAG,Reduction,Horodecki96,HHHH,Peres}
are the positive partial transpose (PPT) and reduction
(RED) criteria, induced by the transpose map $\mathsf T_n$
and the reduction map $R_n$, respectively, defined by
\begin{align*}
\mathsf T_n(A)=A^{\mathsf T},
\quad
R_n(A)=\operatorname{Tr}(A)I_n-A,
\quad A\in\mathrm{Mat}_n(\C).
\end{align*}
We write $K_{\mathrm{PPT}}^{(n,m)}$ and
$K_{\mathrm{RED}}^{(n,m)}$ for the undetected sets
associated with $\mathsf T_n$ and $R_n$, respectively,
and $\mathrm{SEP}^{(n,m)}$ for the set of separable states.
Since every separable state satisfies the PPT criterion,
and the PPT criterion is known to imply the RED criterion, we have
\begin{align*}
\mathrm{SEP}^{(n,m)}
\subseteq K_{\mathrm{PPT}}^{(n,m)}
\subseteq K_{\mathrm{RED}}^{(n,m)}.
\end{align*}

Collins, Hayden, and Nechita \cite{CHN} introduced a family of
maps $\Phi_\mu$ associated with probability distributions $\mu$
through a free-product construction.
When $\Phi_\mu$ is positive, we write
$K_\mu^{(n,m)}=K_{\Phi_\mu}^{(n,m)}$.
In \cite[Section~3.4]{CHN}, the authors studied the corresponding
undetected sets and showed that a small-rank projection family
approaches the reduction criterion in an asymptotic regime.
They also raised the problem of comparing these sets with the PPT
criterion and describing the states that pass all positive free maps.
We study this comparison for freely infinitely divisible distributions.
We show that if $\mu$ is compactly supported and freely
infinitely divisible, and $\Phi_\mu$ is positive, then every
positive operator satisfying the RED criterion also
passes the test associated with $\Phi_\mu$.

\begin{theorem}[Main theorem]\label{thm:main}
Let $n\ge2$, and let $\mu$ be a compactly supported freely
infinitely divisible probability distribution on $\mathbb R$.
If the associated free map
$\Phi_\mu:\mathrm{Mat}_n(\C)\to p\mathcal A p$
defined in \eqref{def:Phi_mu} is positive, then for every $m\ge1$,
\begin{align*}
K_{\mathrm{RED}}^{(n,m)}
\subseteq K_\mu^{(n,m)}.
\end{align*}
\end{theorem}

Together with the standard inclusions above, Theorem~\ref{thm:main} gives
\begin{align*}
\mathrm{SEP}^{(n,m)}
\subseteq K_{\mathrm{PPT}}^{(n,m)}
\subseteq K_{\mathrm{RED}}^{(n,m)}
\subseteq K_\mu^{(n,m)}.
\end{align*}
Thus positive free maps associated with compactly supported freely
infinitely divisible distributions cannot detect PPT entanglement.

The main challenge is to establish the positivity of
$T_\rho=(\Phi_\mu\otimes\id_m)(\rho)$ in the $C^*$-algebra
$p\mathcal A p\otimes\mathrm{Mat}_m(\C)$.
Our strategy is to construct an operator $\widehat T_\rho$
acting on $\mathfrak F(\mathfrak H)\otimes\C^m$
that has the same $*$-distribution as $T_\rho$.
We show that the positivity of $\widehat T_\rho$
implies that of $T_\rho$.
We then derive a sufficient condition for
$\widehat T_\rho\succeq0$ in terms of the positivity
of a $2\times2$ operator-valued matrix.
Finally, we show that this condition holds for every
state satisfying the RED criterion whenever
$\Phi_\mu$ is positive.

The remainder of the paper is organized as follows.
Section~\ref{sec:ncprob} recalls noncommutative probability,
mixed free cumulants, and operators on full Fock spaces.
Section~\ref{sec:construction} introduces the free maps $\Phi_\mu$
and computes the mixed cumulants of their corner variables.
Section~\ref{sec:fid} constructs a Fock realization of these
variables using free L\'evy data and shows that the positivity
of $\Phi_\mu$ implies the positivity of an associated
$2\times2$ operator-valued matrix.
Section~\ref{sec:reduction} establishes a sufficient
condition for the positivity of $\widehat T_\rho$ and proves Theorem~\ref{thm:main}.

%%%%%%%%%%%%%%%%%%%%%%%%%%%%%%%%%%%%%%%%%%%%%%%%%%%%%%%%%%%%%%%%%%%%%%%%%%
%%%%%%%%%%%%%%%%%%%%%%%%%%%%%%%%%%%%%%%%%%%%%%%%%%%%%%%%%%%%%%%%%%%%%%%%%%
%%%
%%% Section 2
%%%
%%%%%%%%%%%%%%%%%%%%%%%%%%%%%%%%%%%%%%%%%%%%%%%%%%%%%%%%%%%%%%%%%%%%%%%%%%
%%%%%%%%%%%%%%%%%%%%%%%%%%%%%%%%%%%%%%%%%%%%%%%%%%%%%%%%%%%%%%%%%%%%%%%%%%
\section{Noncommutative probability and Fock space}\label{sec:ncprob}
\subsection{Freeness and free cumulants}

A \emph{$C^*$-probability space} is a pair
$(\mathcal B,\tau)$ consisting of a unital $C^*$-algebra
$\mathcal B$ and a state $\tau$, that is, a linear functional
satisfying $\tau(1)=1$ and $\tau(b^*b)\ge0$ for every
$b\in\mathcal B$.
The state is \emph{tracial} if $\tau(ab)=\tau(ba)$ for all
$a,b\in\mathcal B$, and \emph{faithful} if
$\tau(b^*b)=0$ implies $b=0$.

Unital $C^*$-subalgebras $(\mathcal B_i)_{i\in I}$ are
\emph{free} with respect to $\tau$ if
\begin{align*}
\tau(a_1\cdots a_r)=0
\end{align*}
whenever $a_j\in\mathcal B_{i_j}$,
$\tau(a_j)=0$, and $i_j\ne i_{j+1}$ for
$1\le j<r$.
Random variables are free when the unital $C^*$-subalgebras
they generate are free.

Free cumulants encode moments through noncrossing partitions.
A partition of $\{1,\ldots,r\}$ is \emph{noncrossing}
if there are no $a<b<c<d$ such that $a,c$ belong to one
block and $b,d$ belong to a different block.
Writing $\NC(r)$ for the set of noncrossing partitions,
the free cumulants are the multilinear functionals
uniquely determined by
\begin{align*}
\tau(a_1\cdots a_r)
&=\sum_{\pi\in\NC(r)}\kappa_\pi(a_1,\ldots,a_r),\\*
\kappa_\pi(a_1,\ldots,a_r)
&=\prod_{\substack{V\in\pi\\V=\{i_1<\cdots<i_k\}}}
\kappa_k(a_{i_1},\ldots,a_{i_k}).
\end{align*}
For a self-adjoint random variable $S$, we abbreviate
$\kappa_\pi(S,\ldots,S)$ to $\kappa_\pi(S)$ for all $\pi \in \NC(r)$.
Mixed free cumulants of free families vanish.

For $S$ free from a unital $C^*$-subalgebra $\mathcal D$
and $b_1,\ldots,b_r\in\mathcal D$, we have
\cite[Theorem~14.4]{NS}
\begin{equation}\label{eq:kreweras}
\tau(Sb_1S b_2\cdots S b_r)
=\sum_{\pi\in\NC(r)}\kappa_\pi(S)
\prod_{V\in K(\pi)}
\tau\left(\prod_{j\in V}^{\longrightarrow}b_j\right).
\end{equation}
The arrow indicates that the factors are ordered by increasing
indices.
Here $K(\pi)$ is the Kreweras complement:
identifying $\pi$ with the permutation whose cycles list
the elements of each block in increasing order,
$K(\pi)$ corresponds to $\pi^{-1}(1\ 2\ \cdots\ r)$.
In particular, $|\pi|+|K(\pi)|=r+1$, where $|\pi|$
denotes the number of blocks.

\subsection{The word model of full Fock space}

Let $\Gamma$ be an orthonormal basis of a complex Hilbert space $\mathfrak H$,
and let $\Gamma^*$ denote the set of finite words
over $\Gamma$, including the empty word $\emptyset$.
We take inner products to be linear in the second variable.
We work with the word realization of the full Fock space
$\mathfrak F(\mathfrak H)$ and its vacuum state
$\omega_\emptyset:\mathcal B(\mathfrak F(\mathfrak H))
\to\mathbb C$, given by
\begin{align*}
\mathfrak F(\mathfrak H)
&=\ell^2(\Gamma^*)
=\mathbb C\emptyset\oplus
\bigoplus_{r\ge1}\ell^2(\Gamma^r),\\*
\omega_\emptyset(A)
&=\langle\emptyset,A\emptyset\rangle.
\end{align*}

For $a\in\Gamma$ and a word $s\in\Gamma^*$, denote their
concatenation by $as$. The left creation operators $\ell(a)$ and the
gauge operators $\Lambda(A)$ are given by
\begin{align*}
\ell(a)s&=as,\quad s\in \Gamma^*,\\*
\Lambda(A)\emptyset&=0,\quad \Lambda(A)(as)=(Aa)s,\quad A\in\mathcal B(\mathfrak H).
\end{align*}
The map $a\mapsto\ell(a)$ extends linearly and continuously
to all of $\mathfrak H$.
The adjoint satisfies $\ell(h)^*\emptyset=0$ and
$\ell(h)^*(as)=\langle h,a\rangle s$.
The pair $(\mathcal B(\Fock(\mathfrak H)),\omega_\emptyset)$ is a
$C^*$-probability space. 

For $h\in \mathfrak H$, $A=A^*\in\mathcal B(\mathfrak H)$ and $\gamma\in\mathbb R$,
consider the self-adjoint random variable
\begin{align*}
J=\gamma I+\ell(h)+\ell(h)^*+\Lambda(A).
\end{align*}
With respect to the vacuum state, all first cumulants of
creation, annihilation, and gauge operators vanish.
Their only possibly nonzero cumulants of order $r\ge2$
have the form
\begin{equation}\label{eq:fock-cumulants}
\kappa_r^{\omega_\emptyset}
\bigl(\ell(\xi)^*,\Lambda(A_2),\ldots,
\Lambda(A_{r-1}),\ell(\eta)\bigr)
=\langle\xi,A_2\cdots A_{r-1}\eta\rangle;
\end{equation}
see \cite[Proposition~13.5]{NS}.
For $r=2$, the product on the right is interpreted as the identity.
Consequently,  the self-adjoint operator
$J$ has free cumulants
\begin{align*}
\kappa_1^{\omega_\emptyset}(J)=\gamma,
\quad
\kappa_r^{\omega_\emptyset}(J)
=\langle h,A^{r-2}h\rangle,\quad r\ge2.
\end{align*}

%%%%%%%%%%%%%%%%%%%%%%%%%%%%%%%%%%%%%%%%%%%%%%%%%%%%%%%%%%%%%%%%%%%%%%%%%%
%%%%%%%%%%%%%%%%%%%%%%%%%%%%%%%%%%%%%%%%%%%%%%%%%%%%%%%%%%%%%%%%%%%%%%%%%%
%%%
%%% Section 3
%%%
%%%%%%%%%%%%%%%%%%%%%%%%%%%%%%%%%%%%%%%%%%%%%%%%%%%%%%%%%%%%%%%%%%%%%%%%%%
%%%%%%%%%%%%%%%%%%%%%%%%%%%%%%%%%%%%%%%%%%%%%%%%%%%%%%%%%%%%%%%%%%%%%%%%%%
\section{The Collins--Hayden--Nechita free-product map}\label{sec:construction}

Let $n\geq2$ and let $(\mathcal B,\tau_\mathcal B)$ be a
$C^*$-probability space with a faithful tracial state, containing
a bounded self-adjoint element $S$ with distribution $\mu$.
Form the reduced tracial $C^*$-free product \cite[Theorem~7.9]{NS}
\begin{align}\label{def:mathcalA}
(\mathcal A,\tau)
=(\mathrm{Mat}_n(\C),\tr_n)*(\mathcal B,\tau_\mathcal B),
\quad \tr_n=n^{-1}\operatorname{Tr}.
\end{align}
Let $\{e_{ij}\}_{i,j=1}^n$ be the standard matrix units of
$\mathrm{Mat}_n(\mathbb C)$, put $p=e_{11}$, and set
$\tau_p=n\tau|_{p\mathcal Ap}$. The reduced free-product state $\tau$ is faithful and tracial.
The corner $p\mathcal Ap$ has unit $p$, and $\tau_p$ is faithful and tracial.
The construction of Collins, Hayden, and Nechita
\cite[Section~3.2]{CHN} associates to the distribution $\mu$ of $S$ the map
\begin{align}\label{def:Phi_mu}
\Phi_\mu:\mathrm{Mat}_n(\C)&\longrightarrow p\mathcal Ap,\\
e_{ij}&\longmapsto x_{ij}:=e_{1i}Se_{j1}. \label{def:corner_fam}
\end{align}
The positivity of the free map $\Phi_\mu$ is characterized
in \cite[Theorem~3.2]{CHN}.
 Write $\kappa_r^p$ for
those computed with respect to $\tau_p$.

We recall the mixed free cumulants of the corner variables $x_{ij}$.
The following is the one-variable case of
\cite[Theorem~14.18]{NS}; we include a proof in our notation.

\begin{lemma}\label{lem:corner}
For any bounded self-adjoint $S$ free from $\mathrm{Mat}_n(\C)$ and for $r\geq1$,
\begin{align*}
\kappa_r^p(x_{i_1j_1},\dots,x_{i_rj_r})
=n^{1-r}\kappa_r(S)\prod_{k=1}^r\delta_{j_k,i_{k+1}},\quad i_{r+1}=i_1.
\end{align*}
\end{lemma}
\begin{proof}
Let $\sigma=(1\ 2\ \cdots\ r)$ and $b_t=e_{j_t,i_{\sigma(t)}}$. Cyclicity and \eqref{eq:kreweras} give the expansion
\begin{align*}
\tau_p(x_{i_1j_1}\cdots x_{i_rj_r})
=n\sum_{\pi\in\NC(r)}\kappa_\pi(S)
\prod_{V\in K(\pi)}\tr_n\left(\prod_{t\in V}^{\longrightarrow}b_t\right).
\end{align*}
Since the cycles of $\theta:=\pi^{-1}\sigma$ are the blocks of $K(\pi)$,
multiplication of matrix units gives
\begin{align*}
\prod_{V\in K(\pi)}
\tr_n\left(\prod_{t\in V}^{\longrightarrow}b_t\right)
&=n^{-|K(\pi)|}
\prod_{V\in K(\pi)}\prod_{t\in V}
\delta_{i_{\sigma(t)},j_{\theta(t)}}\\*
&=n^{-|K(\pi)|}
\prod_{t=1}^r
\delta_{i_{\pi(\theta(t))},j_{\theta(t)}}
=n^{-|K(\pi)|}
\prod_{t=1}^r\delta_{j_t,i_{\pi(t)}}.
\end{align*}
Here we used $\sigma=\pi\theta$ and reindexed the last product by $\theta$.
 The noncrossing identity $|\pi|+|K(\pi)|=r+1$ gives the factor $n^{|\pi|-r}$. Thus
\begin{align*}
\tau_p(x_{i_1j_1}\cdots x_{i_rj_r})
=\sum_{\pi\in\NC(r)}\prod_{V\in\pi}
\left(n^{1-|V|}\kappa_{|V|}(S)
\prod_{t\in V}\delta_{j_t,i_{\pi(t)}}\right).
\end{align*}
This is the noncrossing moment--cumulant formula with the indicated block cumulants. Uniqueness of free cumulants proves the claim.
\end{proof}

%%%%%%%%%%%%%%%%%%%%%%%%%%%%%%%%%%%%%%%%%%%%%%%%%%%%%%%%%%%%%%%%%%%%%%%%%%
%%%%%%%%%%%%%%%%%%%%%%%%%%%%%%%%%%%%%%%%%%%%%%%%%%%%%%%%%%%%%%%%%%%%%%%%%%
%%%
%%% Section 4
%%%
%%%%%%%%%%%%%%%%%%%%%%%%%%%%%%%%%%%%%%%%%%%%%%%%%%%%%%%%%%%%%%%%%%%%%%%%%%
%%%%%%%%%%%%%%%%%%%%%%%%%%%%%%%%%%%%%%%%%%%%%%%%%%%%%%%%%%%%%%%%%%%%%%%%%%

\section{Freely infinitely divisible distributions and their Fock models}\label{sec:fid}
\subsection{Free L\'evy data and joint Fock distribution}

Throughout Sections~\ref{sec:fid} and~\ref{sec:reduction}, fix $n\ge2$
and a compactly supported freely infinitely divisible probability
distribution $\mu$ on $\mathbb R$. Let $S$, $p$, and $\Phi_\mu$ be as in
Section~\ref{sec:construction}.
Free infinite divisibility means that for each integer $k\ge1$, there exists a probability measure $\mu_k$ on $\mathbb R$ such that $\mu=\mu_k^{\boxplus k}$, where $\mu_k^{\boxplus k}$ denotes the $k$-fold free additive convolution of $\mu_k$ with itself.

By the free L\'evy--Khintchine theorem \cite[Theorem~13.16]{NS}, there exist $\gamma\in\mathbb R$ and a compactly supported finite positive measure $\nu$ on $\mathbb R$ such that
\begin{align*}
\kappa_1(\mu)=\gamma,\quad \kappa_r(\mu)=\int_{\mathbb R}t^{r-2}\,d\nu(t), \quad r\geq2.
\end{align*}
Set $\mathfrak H_0=L^2(\nu)$, let $v=1$, and let $X\in\mathcal B(\mathfrak H_0)$ be multiplication by the coordinate function.
Since $\nu$ has compact support, $X$ is bounded and self-adjoint, and
\begin{align}\label{eq:levy-data}
\kappa_1(\mu)=\gamma,\quad
\kappa_r(\mu)=\langle v,X^{r-2}v\rangle,\quad r\ge2.
\end{align}
We call $(\mathfrak H_0,\gamma,v,X)$ the \emph{free L\'evy data} of $\mu$
and retain this notation throughout Sections~\ref{sec:fid}
and~\ref{sec:reduction}. We regard $v$ as the operator
$\C\to\mathfrak H_0$, $z\mapsto zv$, so that
$v^*u=\langle v,u\rangle$.
If $\nu=0$, then $\mathfrak H_0=\{0\}$,
$\mathfrak F(\mathfrak H_0)=\C\emptyset$, and $\mu=\delta_\gamma$.

We now construct a family of operators on the full Fock space over
\begin{align*}
\mathfrak H=\C^{n^2}\otimes\mathfrak H_0
\end{align*}
with the same mixed free cumulants as the corner variables
in \eqref{def:corner_fam}. 
Let $\{f_{ij}:1\le i,j\le n\}$ be an orthonormal basis of
$\C^{n^2}$ and define
\needspace{7\baselineskip}
\begin{align*}
\mathfrak H\ni h_{ij}
&=\frac1{\sqrt n}\,f_{ij}\otimes v,\\
\mathcal B(\mathfrak H)\ni P_{ij}
&=\left(f_{kl}\otimes u\longmapsto
\frac1n\delta_{jk}\,f_{il}\otimes Xu\right).
\end{align*}
Finally, we set 
\begin{align}\label{eq:fock-corner-model}
\widehat x_{ij}
=\gamma\delta_{ij}I_{\mathfrak F(\mathfrak H)}
+\ell(h_{ij})+\ell(h_{ji})^*+\Lambda(P_{ij}) \in \mathcal B(\mathfrak F(\mathfrak H)).
\end{align}
Since $P_{ij}^*=P_{ji}$ and $\Lambda(A)^*=\Lambda(A^*)$, we have $\widehat x_{ij}^*=\widehat x_{ji}$.

\begin{lemma}\label{lem:fock}
The corner variables $x_{ij}$ in $(p\mathcal A p,\tau_p)$ defined in \eqref{def:corner_fam} and the operators $\widehat x_{ij}$ in
$(\mathcal B(\Fock(\mathfrak H)),\omega_\emptyset)$
defined in \eqref{eq:fock-corner-model}
have the same joint $*$-distribution.
\end{lemma}

\begin{proof}
Since $S=S^*$, we have $x_{ij}^*=x_{ji}$.
The Fock realizations satisfy $\widehat x_{ij}^*=\widehat x_{ji}$ as well.
Thus every joint $*$-moment reduces to an ordinary joint moment. By the moment--cumulant formula, it suffices to compare the joint free cumulants.

For $r=1$, we have
\begin{align*}
\kappa_1^{\omega_\emptyset}(\widehat x_{ij})
=\gamma\delta_{ij}
=\kappa_1^p(x_{ij}).
\end{align*}

Let $r\ge2$. Identity terms do not contribute to free cumulants of order at least two.
By multilinearity and \eqref{eq:fock-cumulants}, the only nonzero contribution gives
\begin{align*}
\kappa_r^{\omega_\emptyset}
(\widehat x_{i_1j_1},\ldots,
 \widehat x_{i_rj_r})&=
\left\langle h_{j_1i_1},
P_{i_2j_2}\cdots P_{i_{r-1}j_{r-1}}
h_{i_rj_r}\right\rangle\\*
&=
n^{1-r}\langle v,X^{r-2}v\rangle
\prod_{k=1}^r\delta_{j_k,i_{k+1}},
\quad i_{r+1}=i_1.
\end{align*}
By \eqref{eq:levy-data} and Lemma~\ref{lem:corner}, this equals
$
\kappa_r^p(x_{i_1j_1},\ldots,x_{i_rj_r})
$.
Hence the two families have the same joint $*$-distribution.
\end{proof}

\begin{corollary}
\label{cor:positivity-transfer}
Let $m\geq1$, and let $\rho=[\rho_{ij}]$ be self-adjoint in $\mathrm{Mat}_n(\mathrm{Mat}_m(\mathbb C))$. Set
\begin{align*}
T_\rho
&=\sum_{i,j}x_{ij}\otimes\rho_{ij}
\in(p\mathcal A p)\otimes \mathrm{Mat}_m(\mathbb C),\\*
\widehat T_\rho
&=\sum_{i,j}\widehat x_{ij}\otimes\rho_{ij}
\in\mathcal B(\Fock(\mathfrak H))\otimes \mathrm{Mat}_m(\mathbb C).
\end{align*}
If $\widehat T_\rho\succeq0$, then $T_\rho\succeq0$.
\end{corollary}

\begin{proof}
Recall that $\tau_p$ is a faithful tracial state on $p\mathcal Ap$. 
Hence $\tau_p\otimes\tr_m$ is faithful.

By Lemma~\ref{lem:fock}, for every $r\ge1$,
\begin{align*}
(\tau_p\otimes\tr_m)(T_\rho^r)&=\sum_{i_1,j_1,\ldots,i_r,j_r}
\tau_p(x_{i_1j_1}\cdots x_{i_rj_r})
\tr_m(\rho_{i_1j_1}\cdots\rho_{i_rj_r})\\*
&=\sum_{i_1,j_1,\ldots,i_r,j_r}
\omega_\emptyset(
\widehat x_{i_1j_1}\cdots\widehat x_{i_rj_r})
\tr_m(\rho_{i_1j_1}\cdots\rho_{i_rj_r})\\*
&=(\omega_\emptyset\otimes\tr_m)
(\widehat T_\rho^r).
\end{align*}
Since both operators are bounded and self-adjoint, their spectral measures are compactly supported and determined by their moments.
Thus $\widehat T_\rho\succeq0$ implies that the spectral measure of $T_\rho$ is supported on $[0,\infty)$.
By faithfulness of $\tau_p\otimes\tr_m$, this support equals $\operatorname{spec}(T_\rho)$.
Hence $T_\rho\succeq0$.
\end{proof}

\subsection{A positivity certificate from the Fock model}\label{sec:scalar-certificate}
We derive a scalar inequality from the free L\'evy data that will
later serve as a positivity certificate for $\widehat T_\rho$.

\begin{lemma}\label{lem:ball-dual}
For the free L\'evy data $(\mathfrak H_0,\gamma,v,X)$ of $\mu$,
the following are equivalent:
\begin{enumerate}
\item[(i)] For every $u\in\mathfrak H_0$ with $\|u\|\le1$,
\begin{align*}
\gamma+\frac{2}{\sqrt n}\operatorname{Re}\langle v,u\rangle
+\frac1n\langle u,Xu\rangle\ge0.
\end{align*}
\item[(ii)] There exists $q\ge0$ such that
\begin{align*}
\begin{pmatrix}
\gamma-q&v^*/\sqrt n\\
v/\sqrt n&X/n+qI_{\mathfrak H_0}
\end{pmatrix}\succeq0
\quad\text{on }\C\oplus\mathfrak H_0.
\end{align*}
\end{enumerate}
\end{lemma}

\begin{proof}
For $q\ge0$, put
\begin{align*}
f_q(u)
=\gamma-q+\frac{2}{\sqrt n}\operatorname{Re}
\langle v,u\rangle
+\left\langle u,
\left(\frac Xn+qI_{\mathfrak H_0}\right)u
\right\rangle.
\end{align*}
The positivity in (ii) is equivalent to
$f_q(u)\ge0$ for every $u\in\mathfrak H_0$.
Indeed, evaluating its quadratic form at
$(1,u)^{\mathsf T}$ gives $f_q(u)$;
conversely, homogeneity gives nonnegativity
at every $(z,u)^{\mathsf T}$ with $z\ne0$, and
continuity covers $z=0$.

If (ii) holds, then for $\|u\|\le1$,
\begin{align*}
\gamma+\frac{2}{\sqrt n}
\operatorname{Re}\langle v,u\rangle
+\frac1n\langle u,Xu\rangle=f_q(u)+q(1-\|u\|^2)\ge0.
\end{align*}
Thus (ii) implies (i).

Conversely, assume (i), and write
\begin{align*}
F(u)=\gamma+\frac{2}{\sqrt n}
\operatorname{Re}\langle v,u\rangle
+\frac1n\langle u,Xu\rangle.
\end{align*}
Taking $u=0$ gives $\gamma\ge0$.
If $\mathfrak H_0=\{0\}$, then (ii)
holds with $q=0$.

Suppose $\mathfrak H_0\ne\{0\}$.
Let $E\subseteq\mathfrak H_0$ be a
finite-dimensional complex subspace,
and let $P_E$ denote the orthogonal
projection onto $E$. Set
\begin{align*}
A_E=\frac1nP_EX|_E,
\quad
b_E=\frac1{\sqrt n}P_Ev.
\end{align*}
Then $A_E=A_E^*$, and for every $u\in E$,
\begin{align*}
F(u)=\gamma+2\operatorname{Re}\langle b_E,u\rangle
+\langle u,A_Eu\rangle.
\end{align*}

Apply the complex extended S-lemma
\cite[Theorem~2.1]{BeckEldar}
to $F|_E$, $1-\|u\|^2$, and the constant
function $1$. Since $u=0$ is strictly
feasible, there exist $q_E,\beta_E\ge0$
such that
\begin{align*}
\begin{pmatrix}
A_E&b_E\\*
b_E^*&\gamma
\end{pmatrix}
\succeq
q_E\begin{pmatrix}
-I_E&0\\*
0&1
\end{pmatrix}
+\beta_E\begin{pmatrix}
0&0\\*
0&1
\end{pmatrix}.
\end{align*}
Consequently,
\begin{align*}
M_{q_E,E}\succeq0,
\quad\text{where}\quad
M_{q,E}:=\begin{pmatrix}
\gamma-q&b_E^*\\*
b_E&A_E+qI_E
\end{pmatrix}.
\end{align*}
In particular, $0\le q_E\le\gamma$.

For each finite-dimensional subspace
$E\subseteq\mathfrak H_0$, define
$
Q_E=\left\{
q\in[0,\gamma]:
M_{q,E}\succeq0
\right\}
$.
Each $Q_E$ is nonempty and closed,
since the matrix depends continuously
on $q$. Moreover, for finite-dimensional
subspaces $E_1,\ldots,E_k$,
\begin{align*}
Q_{E_1+\cdots+E_k}
\subseteq\bigcap_{j=1}^kQ_{E_j}.
\end{align*}
Thus the family $\{Q_E\}$ has the
finite-intersection property.
By compactness of $[0,\gamma]$,
there exists $
q\in\bigcap_E Q_E
$.

For arbitrary $z\in\C$ and $u\in\mathfrak H_0$,
choose a finite-dimensional subspace $E$ containing $u$.
Since $P_Eu=u$ and $q\in Q_E$, we have
\begin{align*}
\left\langle
\begin{pmatrix}z\\u\end{pmatrix},
M_q\begin{pmatrix}z\\u\end{pmatrix}
\right\rangle=\left\langle
\begin{pmatrix}z\\u\end{pmatrix},
M_{q,E}
\begin{pmatrix}z\\u\end{pmatrix}
\right\rangle
\ge0.
\end{align*}
Hence $M_q\succeq0$, proving (ii).
\end{proof}

We next transfer positivity of the corner variable $pSp$ to its
Fock realization by verifying faithfulness on the generated algebra.

\begin{lemma}\label{lem:vacuum-detection}
For the free L\'evy data $(\mathfrak H_0,\gamma,v,X)$ of $\mu$, set
\begin{align*}
J_n=\gamma I+\ell(v/\sqrt n)+\ell(v/\sqrt n)^*
+\Lambda(X/n)\in \mathcal B(\Fock(\mathfrak H_0)).
\end{align*}
If $pSp\succeq0$, then $J_n\succeq0$.
\end{lemma}

\begin{proof}
If $\nu=0$, then $\mu=\delta_\gamma$ and faithfulness gives
$S=\gamma1$. Hence $pSp=\gamma p\succeq0$ implies
$\gamma\ge0$, and $J_n=\gamma I\succeq0$ on
$\mathfrak F(\{0\})=\C\emptyset$.

Suppose now that $\nu\ne0$. Recall that
\begin{align*}
\mathfrak H_0=L^2(\nu),\quad
w=\frac1{\sqrt n},\quad
(Bf)(t)=\frac tn f(t),
\end{align*}
and
\begin{align*}
J_n=\gamma I+\ell(w)+\ell(w)^*+\Lambda(B).
\end{align*}

We apply the vacuum faithfulness result of
Yang~\cite[Lemma~3.5]{Yang}.
To verify its hypotheses, consider the dense
subspace
\begin{align*}
\mathscr D=L^\infty(\nu)\subseteq L^2(\nu),
\end{align*}
equipped with the product and involution
\begin{align*}
(f\star g)(t)
=\frac{t}{\sqrt n}f(t)g(t),
\quad
f^\sharp(t)=\overline{f(t)}.
\end{align*}
Since $\nu$ has compact support,
$\mathscr D$ is an associative involutive
algebra.

For every $f\in\mathscr D$, left multiplication
by $f$ is bounded, since
\begin{align*}
\|f\star g\|_2
&\le
\left\|\frac{tf}{\sqrt n}\right\|_\infty
\|g\|_2.
\end{align*}
Moreover, for $f,g,h\in\mathscr D$,
\begin{align*}
\langle f\star g,h\rangle
&=\int_{\mathbb R}
\frac{t}{\sqrt n}
\overline{f(t)g(t)}h(t)\,d\nu(t)=\langle g,f^\sharp\star h\rangle.
\end{align*}
Finally, the involution is closable.
Indeed, it satisfies
$
\|f^\sharp\|_2=\|f\|_2
$.
Thus, if $f_k\to0$ and $f_k^\sharp\to g$
in $L^2(\nu)$, then
\begin{align*}
\|g\|_2
=\lim_{k\to\infty}\|f_k^\sharp\|_2
=\lim_{k\to\infty}\|f_k\|_2
=0.
\end{align*}
Consequently, $\mathscr D$ is a pseudo left
Hilbert algebra in the sense of
Yang~\cite[Definition~2.6]{Yang}.

Let $\pi_l$ denote its left multiplication
representation. 
Since $w^\sharp=w$ and $\pi_l(w)=B$,
the Fock operator associated with $w$ in
Yang's construction is
\begin{align*}
\ell(w)+\ell(w^\sharp)^*+\Lambda(\pi_l(w))
=\ell(w)+\ell(w)^*+\Lambda(B)
=J_n-\gamma I.
\end{align*}
Consequently,
\begin{align*}
C^*(J_n)\subseteq\Gamma(\mathscr D),
\end{align*}
where $\Gamma(\mathscr D)$ is the von Neumann
algebra generated by the Fock operators
associated with $\mathscr D$.

By Yang's Lemma~3.5, the vacuum vector
$\emptyset$ is separating for
$\Gamma(\mathscr D)$. In particular, the
vacuum state $\omega_{\emptyset}$ is faithful
on $C^*(J_n)$.

On the other hand, Lemma~\ref{lem:corner} and
\eqref{eq:fock-cumulants} give
\begin{align*}
\kappa_r^{\omega_\emptyset}(J_n)
=n^{1-r}\kappa_r(\mu)
=\kappa_r^p(pSp),\quad r\ge1.
\end{align*}
Thus $J_n$ and $pSp$ have the same moments under
$\omega_{\emptyset}$ and $\tau_p$, respectively. Since both
operators are bounded and self-adjoint,
their continuous functional calculi have
the same expectations.
Set
\begin{align*}
T=(J_n)_-
=\frac{|J_n|-J_n}{2},
\quad
|J_n|=(J_n^2)^{1/2}.
\end{align*}
Then $T\in C^*(J_n)$ and
\begin{align*}
\|T\emptyset\|^2=\omega_{\emptyset}(T^2)=\tau_p\bigl(((pSp)_-)^2\bigr)=0,
\end{align*}
where the last equality follows from
$pSp\succeq0$.

Since $\emptyset$ is separating for
$C^*(J_n)$, it follows that $T=0$.
Therefore $J_n\succeq0$.
\end{proof}

Testing the positive Fock operator on tensor-power vectors now
produces the inequality in Lemma~\ref{lem:ball-dual}.

\begin{proposition}\label{prop:scalar-certificate}
For the free L\'evy data $(\mathfrak H_0,\gamma,v,X)$ of $\mu$,
if $\Phi_\mu$ is positive,
then there exists $q\ge0$, depending only on $\mu$ and $n$, such that
\begin{equation}\label{eq:scalar-certificate}
\begin{pmatrix}
\gamma-q&v^*/\sqrt n\\
v/\sqrt n&X/n+qI_{\mathfrak H_0}
\end{pmatrix}\succeq0.
\end{equation}
In particular, $q$ is independent of the ancillary dimension
and of the input state.
\end{proposition}

\begin{proof}
Since $e_{11}\succeq0$, positivity of $\Phi_\mu$
gives
\begin{align*}
pSp=\Phi_\mu(e_{11})\succeq0.
\end{align*}
Put $w=v/\sqrt n$
and $B=X/n$. By Lemma~\ref{lem:vacuum-detection},
we have $J_n\succeq0$.

For $u\in\mathfrak H_0$ with $\|u\|<1$, define
\begin{align*}
\xi_u=\sum_{r\ge0}u^{\otimes r}
\in\Fock(\mathfrak H_0),
\quad u^{\otimes0}=\emptyset.
\end{align*}
By the definitions of the creation and gauge
operators,
\begin{align*}
\|\xi_u\|^2=\frac1{1-\|u\|^2},\quad
\langle\xi_u,\ell(w)\xi_u\rangle=\frac{\langle u,w\rangle}{1-\|u\|^2},\quad
\langle\xi_u,\Lambda(B)\xi_u\rangle=\frac{\langle u,Bu\rangle}{1-\|u\|^2}.
\end{align*}
Consequently, $J_n\succeq0$ implies
\begin{align*}
0&\le(1-\|u\|^2)
\langle\xi_u,J_n\xi_u\rangle\\*
&=\gamma+2\operatorname{Re}\langle w,u\rangle
+\langle u,Bu\rangle\\*
&=\gamma+\frac2{\sqrt n}
\operatorname{Re}\langle v,u\rangle
+\frac1n\langle u,Xu\rangle.
\end{align*}
By continuity, this inequality also holds
when $\|u\|=1$.
Lemma~\ref{lem:ball-dual} therefore gives
\eqref{eq:scalar-certificate}.
\end{proof}

%%%%%%%%%%%%%%%%%%%%%%%%%%%%%%%%%%%%%%%%%%%%%%%%%%%%%%%%%%%%%%%%%%%%%%%%%%
%%%%%%%%%%%%%%%%%%%%%%%%%%%%%%%%%%%%%%%%%%%%%%%%%%%%%%%%%%%%%%%%%%%%%%%%%%
%%%
%%% Section 5
%%%
%%%%%%%%%%%%%%%%%%%%%%%%%%%%%%%%%%%%%%%%%%%%%%%%%%%%%%%%%%%%%%%%%%%%%%%%%%
%%%%%%%%%%%%%%%%%%%%%%%%%%%%%%%%%%%%%%%%%%%%%%%%%%%%%%%%%%%%%%%%%%%%%%%%%%

\section{Proof of the reduction bound}\label{sec:reduction}

\subsection{A positivity criterion for the Fock realization}\label{sec:fock_pos_cri}
Recall that we construct the space $\mathfrak H=\C^{n^2}\otimes\mathfrak H_0$ from the free L\'evy data $(\mathfrak H_0,\gamma,v,X)$.
Define the bounded operators
\begin{align*}
C_\rho:\C^m&\longrightarrow\mathfrak H\otimes\C^m,\\*
u &\longmapsto\sum_{i,j}h_{ij}\otimes\rho_{ij}u,\\*
P_\rho&=\sum_{i,j}P_{ij}\otimes\rho_{ij}
\in\mathcal B(\mathfrak H\otimes\C^m).
\end{align*}
Since $\rho_{ij}^*=\rho_{ji}$ and
$P_{ij}^*=P_{ji}$, we have $P_\rho=P_\rho^*$.

\begin{lemma}\label{lem:fock-assembly}
Let $\rho=[\rho_{ij}]\in\mathrm{Mat}_n(\mathrm{Mat}_m(\C))$
be self-adjoint and $D_\rho=\sum_{i}\rho_{ii}$. If
\begin{align*}
B_{\rho,q}=
\begin{pmatrix}
\gamma D_\rho -qI_m&C_\rho^*\\*
C_\rho&P_\rho+qI
\end{pmatrix}\succeq0
\end{align*}
for some $q\ge0$, then
$\widehat T_\rho\succeq0$.
\end{lemma}

\begin{proof}
Choose an orthonormal basis $\Gamma$ of $\mathfrak H$.
For a finite-word vector
$\xi=\sum_{s\in\Gamma^*}s\otimes z_s$, where 
$ z_s\in\C^m$,
put
\begin{align*}
b_s=\sum_{a\in\Gamma}a\otimes z_{as}
\in\mathfrak H\otimes\C^m.
\end{align*}
Since every nonempty word has a unique first letter and suffix,
\begin{align}\label{eq:assembly-word-norm}
\sum_{s\in\Gamma^*}\|b_s\|^2=\sum_{s\in\Gamma^*}\sum_{a\in\Gamma}
\|z_{as}\|^2=\sum_{\substack{t\in\Gamma^*\\t\ne\emptyset}}
\|z_t\|^2=\|\xi\|^2-\|z_\emptyset\|^2.
\end{align}

We compute the contributions of the scalar, creation,
annihilation, and gauge operators separately.

For the scalar term,
\begin{align*}
\left\langle\xi,
\left(\gamma I\otimes D_\rho\right)\xi
\right\rangle
=\gamma\sum_s\langle z_s,D_\rho z_s\rangle.
\end{align*}
For the creation term,
\begin{align*}
\left\langle\xi,
\left(\sum_{i,j}\ell(h_{ij})\otimes\rho_{ij}\right)
\xi\right\rangle&=\sum_{i,j}\sum_{s,t\in \Gamma^*}
\left\langle t\otimes z_t,
(h_{ij}s)\otimes\rho_{ij}z_s\right\rangle\\*
&=\sum_{i,j}\sum_{s\in\Gamma^*}\sum_{a\in\Gamma}
\left\langle a\otimes z_{as},
h_{ij}\otimes\rho_{ij}z_s\right\rangle\\*
&=\sum_{s\in\Gamma^*}\langle b_s,C_\rho z_s\rangle.
\end{align*}
Here the second equality follows because only
words of the form $t=as$ contribute.
Since the annihilation term is the adjoint of
the creation term, their sum contributes
\begin{align*}
2\operatorname{Re}\sum_s
\langle b_s,C_\rho z_s\rangle.
\end{align*}
For the gauge term,
\begin{align*}
\left\langle\xi,
\left(\sum_{i,j}\Lambda(P_{ij})\otimes\rho_{ij}\right)
\xi\right\rangle&=\sum_{i,j}\sum_s\sum_{a,b\in\Gamma}
\left\langle a\otimes z_{as},
(P_{ij}b)\otimes\rho_{ij}z_{bs}\right\rangle\\*
&=\sum_s\langle b_s,P_\rho b_s\rangle.
\end{align*}
Here the gauge operators act on the first
letter while preserving the suffix.
Combining these contributions yields
\begin{align*}
\langle\xi,\widehat T_\rho\xi\rangle
=\gamma\sum_s\langle z_s,D_\rho z_s\rangle
+2\operatorname{Re}\sum_s\langle b_s,C_\rho z_s\rangle+\sum_s\langle b_s,P_\rho b_s\rangle.
\end{align*}

On the other hand, expanding the block quadratic forms gives
\begin{align*}
&\sum_s\left\langle
\begin{pmatrix}z_s\\b_s\end{pmatrix},
B_{\rho,q}
\begin{pmatrix}z_s\\b_s\end{pmatrix}
\right\rangle\\*
=&\gamma\sum_s\langle z_s,D_\rho z_s\rangle
+2\operatorname{Re}
\sum_s\langle b_s,C_\rho z_s\rangle+\sum_s\langle b_s,P_\rho b_s\rangle
-q\sum_s\|z_s\|^2
+q\sum_s\|b_s\|^2\\*
=&\langle\xi,\widehat T_\rho\xi\rangle
-q\|z_\emptyset\|^2,
\end{align*}
where the last equality follows from
\eqref{eq:assembly-word-norm}.

Consequently,
\begin{align}\label{eq:assembly-identity}
\langle\xi,\widehat T_\rho\xi\rangle
={}&\sum_s\left\langle
\begin{pmatrix}z_s\\b_s\end{pmatrix},
B_{\rho,q}
\begin{pmatrix}z_s\\b_s\end{pmatrix}
\right\rangle
+q\|z_\emptyset\|^2.
\end{align}
If $B_{\rho,q}\succeq0$ and $q\ge0$,
the right-hand side is nonnegative.
Since finite-word vectors are dense in
$\Fock(\mathfrak H)\otimes\C^m$ and
$\widehat T_\rho$ is bounded and self-adjoint,
we conclude that $\widehat T_\rho\succeq0$.
\end{proof}

\subsection{Proof of the main theorem}\label{sec:application}

\begin{proof}[Proof of Theorem~\ref{thm:main}]
Fix $\rho\in K_{\mathrm{RED}}^{(n,m)}$ and first suppose
$D_\rho\succ0$. Put
\begin{align*}
R=(I_n\otimes D_\rho^{-1/2})\rho
  (I_n\otimes D_\rho^{-1/2}).
\end{align*}
Since $\rho\succeq0$, we have $R\succeq0$.
The reduction condition gives
\begin{align*}
I_{nm}-R
=(I_n\otimes D_\rho^{-1/2})
 (I_n\otimes D_\rho-\rho)
 (I_n\otimes D_\rho^{-1/2})\succeq0.
\end{align*}
Writing $R=[R_{ij}]\in\mathrm{Mat}_n(\mathrm{Mat}_m(\C))$,
we also have
\begin{align*}
\sum_{i=1}^n R_{ii}
=D_\rho^{-1/2}\left(\sum_{i=1}^n\rho_{ii}\right)
 D_\rho^{-1/2}=I_m.
\end{align*}

Define
\begin{align*}
\widetilde R
=\bigl[\delta_{jl}R_{ik}\bigr]_{(i,j),(k,l)}
\in\mathrm{Mat}_{n^2}(\mathrm{Mat}_m(\C)).
\end{align*}
Let $\{e_i\}_{i=1}^n$ and $\{f_{ij}\}_{i,j=1}^n$
be the standard bases of $\C^n$ and $\C^{n^2}$.
The unitary
\begin{align*}
U:\C^n\otimes\C^m\otimes\C^n
&\longrightarrow\C^{n^2}\otimes\C^m,\\*
e_i\otimes z\otimes e_j&\longmapsto f_{ij}\otimes z
\end{align*}
implies that  $\widetilde R=U(R\otimes I_n)U^*$ and
$
0\preceq\widetilde R\preceq I_{n^2m}
$.
For $z\in\C^m$, set
$
d_z=\sum_{k=1}^n f_{kk}\otimes z
$.
Then
\begin{align}\label{eq:diagonal-compression}
\widetilde R d_z
&=\sum_{i,k}f_{ik}\otimes R_{ik}z
\end{align}
and $
\langle d_z,\widetilde R d_z\rangle=\left\langle z,\sum_kR_{kk}z\right\rangle
=\|z\|^2$.
To fix the tensor-factor order, define the unitary
\begin{align*}W:(\C^{n^2}\otimes\C^m)\otimes\mathfrak H_0
&\longrightarrow \mathfrak H\otimes\C^m,\notag\\*
f_{ij}\otimes z\otimes u
&\longmapsto f_{ij}\otimes u\otimes z.
\end{align*}

The coefficient operators from Section~\ref{sec:fock_pos_cri}
satisfy the identities
\begin{align}\label{eq:CR-PR}
C_Rz&=W\left((\widetilde R d_z)\otimes\frac{v}{\sqrt n}\right),\\
P_R&=W\left(\widetilde R\otimes\frac Xn\right)W^*.\label{eq:CR-PR2}
\end{align}

Let $q\ge0$ be given by Proposition~\ref{prop:scalar-certificate}
and write
\begin{align*}
M_q=
\begin{pmatrix}
\gamma-q&v^*/\sqrt n\\*
v/\sqrt n&X/n+qI_{\mathfrak H_0}
\end{pmatrix}\succeq0.
\end{align*}
Set $E=\C^{n^2}\otimes\C^m$. For
$z\in\C^m$ and $\eta\in\mathfrak H\otimes\C^m$, put
\begin{align*}
\eta'&=W^*\eta\in E\otimes\mathfrak H_0,\\*
\zeta&=\begin{pmatrix}
\widetilde R^{1/2}d_z\\*
(\widetilde R^{1/2}\otimes I_{\mathfrak H_0})\eta'
\end{pmatrix}
\in E\oplus (E\otimes \mathfrak H_0),
\end{align*}
where we identify $E\oplus(E\otimes\mathfrak H_0)$ with
$E\otimes(\C\oplus\mathfrak H_0)$ via
$(a,0)\mapsto a\otimes(1,0)$ and
$(0,b\otimes u)\mapsto b\otimes(0,u)$.
Using \eqref{eq:diagonal-compression}--\eqref{eq:CR-PR2}, we obtain
\begin{align*}
\langle\zeta,(I_E\otimes M_q)\zeta\rangle
=(\gamma-q)\|z\|^2
+2\operatorname{Re}\langle\eta,C_Rz\rangle+\langle\eta,P_R\eta\rangle
+q\langle\eta',(\widetilde R\otimes I_{\mathfrak H_0})\eta'\rangle.
\end{align*}
Consequently,
\begin{align}\label{eq:lifting-decomposition}
\left\langle
\begin{pmatrix}z\\\eta\end{pmatrix},
B_{R,q}\begin{pmatrix}z\\\eta\end{pmatrix}
\right\rangle=\langle\zeta,(I_E\otimes M_q)\zeta\rangle
+q\left\langle\eta',
\bigl((I_E-\widetilde R)\otimes I_{\mathfrak H_0}\bigr)
\eta'\right\rangle.
\end{align}
Both terms on the right are nonnegative, since
$M_q\succeq0$, $q\ge0$, and $\widetilde R\preceq I_E$.
Hence $B_{R,q}\succeq0$.

Lemma~\ref{lem:fock-assembly} gives
$\widehat T_R\succeq0$, and
Corollary~\ref{cor:positivity-transfer} implies
$T_R\succeq0$. Undoing the normalization,
\begin{align*}
T_\rho
=(p\otimes D_\rho^{1/2})T_R
 (p\otimes D_\rho^{1/2})\succeq0.
\end{align*}

If $D_\rho$ is singular, let $e$ be the orthogonal
projection onto $\operatorname{Ran}(D_\rho)$.
For $u\in\ker D_\rho$, positivity of each diagonal
block gives
\begin{align*}
0=\langle u,D_\rho u\rangle
=\sum_i\langle u,\rho_{ii}u\rangle
\quad\Longrightarrow\quad\rho_{ii}u=0.
\end{align*}
Since $\rho\succeq0$, this implies
$\rho(e_j\otimes u)=0$ for each $j$, so
$\rho_{ij}(I_m-e)=0$ for every $i,j$.
Taking adjoints and exchanging $i,j$ gives
$(I_m-e)\rho_{ij}=0$. Consequently,
\begin{align*}
\rho_{ij}=e\rho_{ij}e
\quad (1\le i,j\le n).
\end{align*}
Since $D_\rho$ is positive definite on $e\C^m$,
the preceding argument applies on that space.
As $\rho$ vanishes on its orthogonal complement,
we obtain $T_\rho\succeq0$.
Therefore
\begin{align*}
K_{\mathrm{RED}}^{(n,m)}
\subseteq K_\mu^{(n,m)}.
\end{align*}
\end{proof}

\end{document}